\documentclass{aa}
\usepackage{graphicx}
\usepackage{hyperref}
\usepackage{txfonts}

\newcommand{\meantau}{\overline{\tau}(t_i)}

\newcommand{\tmax}{t_\mathrm{SSN\_{max}}}  

\newcommand{\tmin}{t_\mathrm{SSN\_{min}}}
\newcommand{\id}{{\rm d}}

\begin{document} 

   \title{Detecting stellar activity cycles in p-mode travel times}

   \subtitle{Proof of concept using SOHO/VIRGO  solar observations}

   \author{V.~Vasilyev\inst{1}
          \and
          L.~Gizon\inst{1,2,3}
          }

   \institute{
   Max-Planck-Institut f\"ur Sonnensystemforschung,  G\"ottingen, Germany          
         \and
     Institut für Astrophysik, Georg-August-Universit\"at G\"ottingen,  G\"ottingen, Germany 
        \and 
    Center for Space Science, NYUAD Institute, New York University Abu~Dhabi, Abu~Dhabi, UAE 
}

   \date{Received XXXXX 2023; accepted XXXXX 2023}

\abstract
{
The 11 year solar cycle is known to affect the global modes of solar acoustic oscillations. In particular, p mode frequencies  increase with solar activity.
}{We propose a new method to detect the solar cycle from the  p-mode autocorrelation function, and we validate this method using VIRGO/SPM photometric time series from solar cycles 23 and 24.
} 
{
The p-mode autocorrelation function shows multiple wavepackets separated by time lags of $\sim 123$~min. Using a one-parameter fitting method (from local helioseismology), we measure the seismic travel times from each wavepacket up to skip number 40.
}
{We find that the travel-time variations due to the solar cycle strongly depend on the skip number, with the strongest signature in odd skips from 17 to 31.
Taking the noise covariance into account, the travel-time perturbations can be averaged over all skip numbers to enhance the signal-to-noise ratio. 
}{This method is robust to noise, simpler to implement than peak bagging in the frequency domain, and is promising for asteroseismology. We estimate that the activity cycle of a Sun-like star should be detectable with this new method in {\it Kepler}-like observations down to a visual magnitude of $m_K \sim 11$. However, for fainter stars,  activity cycles are easier to detect in the photometric variability on rotational timescales.
}

   \keywords{solar activity, stellar activity, helioseismology, asteroseismology}
   \maketitle
%


\section{Introduction}
In the case of the Sun, magnetic activity follows an 11 year cycle, which is seen in many observables on the solar surface and in the  atmosphere.  This cycle is prominent in the variations of the sunspot number and the sunspot area, which are used as the standard proxies of solar activity. 
 In the chromosphere, solar active regions lead to increased emission, which is most evident in the core of the CaII H\&K lines  \citep[e.g.,][]{1998ASPC..140..293W}. 
Activity cycles  on Sun-like stars  produce  both long-term spectral and photometric variations detected with ground-based spectroscopic  (see e.g. Mount Wilson Observatory HK Project \citep{Wilson1978, MountWilson1991, MountWilson1995} and photometric observations \citep[see e.g.,][]{Radick_2018}.    High-precision space photometry made it possible to find evidence for activity cycles in thousands of stars \citep{Reinhold2017}.

Methods of helio- and asteroseismology provide an independent way to study magnetic activity. The sunspot cycle is known to affect solar p modes, as seen in the variations of their frequencies \citep[see e.g.,][]{1985Natur.318..449W, 1987A&A...177L..47F, Jimenez-Reyes1998, Chaplin2003, Salabert2004, Howe2018},  line widths, amplitudes \citep[see e.g.,][]{1990LNP...367..129P, 1992A&A...255..363A, 2007A&A...463.1181S}, and energy supply rates \citep{Kiefer2021}.
Mode frequencies and line widths increase with activity, and mode amplitudes decrease
 \cite[see e.g.][for a review]{2016LRSP...13....2B}.
In particular, the large frequency separation is affected \citep{Broomhall2011}.
Modes with higher frequencies are more affected, from which we deduce that   magnetic perturbations act on the modes in the surface layers \citep{Libbrecht1990}.
 \cite{Santos2016} find that sunspots contribute around 30\% of the frequency shifts of low-degree p modes.  
 A contribution from solar activity at high latitudes is not excluded
 \citep{Moreno_Insertis_Solanki_2000}.
 
The first detection of activity using stellar p modes was reported for the F5V CoRoT star HD49933 by \cite{2010Sci...329.1032G}. 
A modulation of the mode frequencies and  amplitudes over 120 days was detected, with higher frequencies corresponding to smaller amplitudes. In addition, frequency shifts are larger at higher frequencies, indicating a near-surface  effect \citep{2011A&A...530A.127S}. Using \textit{Kepler} high-cadence (1 min) data, \cite{Salabert2018} and 
\cite{Santos2019} extended this analysis to a larger sample of Sun-like  stars. Frequency shifts decrease with stellar age and surface rotation period and correlate with effective temperature.

Measuring the frequency shifts of individual modes ---which are often much lower than  1~$\mu$Hz ----  is very challenging, especially at low signal-to-noise ratios.  \cite{Pale1989} proposed to use a cross-correlation technique to measure a mean frequency shift over all modes in the power spectrum. 
\cite{Regulo2016} and \cite{Kiefer2017}  applied the cross-correlation method to analyze $Kepler$ stars. In more than half the stars, periodic variations of the mean frequency shifts were accompanied by variations in other activity proxies, such as   the photometric activity \citep{Santos2018}.

Fitting oscillation power spectra is a delicate enterprise that requires  many parameters:  several parameters for each mode and parameters for the background noise. 
Here, we propose to extract  seismic information from the autocovariance of the intensity time series, $I$, over a segment of the data of length  $T$:
\begin{equation}
    C(t) = \int_0^T I(t') I(t'+t)\ \id t' .
\end{equation} The function $C$ displays a series of wavepackets that are nearly regularly spaced in time, each associated with a particular  arrival time (or skip number). For each skip number, we extract a single parameter: the travel-time perturbation between  $C$  and the smooth reference $C_{\rm ref}$. This one-parameter fit ---first developed in time--distance helioseismology \citep{Gizon2002}--- has been shown to be very robust to noise \cite {Gizon2004} and is very easy to implement (Section~\ref{sec:TTVmethod}). It has become a standard method in this field and has been used, for example, to infer  the Sun's meridional flow  \citep{Gizon2020Science}.  

For each wavepacket, the travel-time perturbation $\tau$ is extracted from $C(t)$ by cross-correlation with a sliding reference $C_{\rm ref}(t-\tau)$.
The function $C_\mathrm{ref}$ is an estimate of the  expectation value of $C$,  obtained from a model or from an average over many realizations  of $C$.
Because the function $C$ is noisy, the travel-time perturbation measured by cross-correlation with $C_{\rm ref}$ is also noisy.
To handle this problem, the  idea is to replace $C$ by the smooth function $C_\epsilon = \epsilon C + (1-\epsilon) C_{\rm ref}$, where $\epsilon$ is small. The travel-time perturbation $\epsilon\tau$ can now unambiguously be measured from $C_\epsilon$ by comparison with the sliding reference $C_{\rm ref}$. Formally, this procedure  can be summarized as follows:
\begin{equation}
\tau    = \lim_{\epsilon \rightarrow 0} \left\{ \frac{1}{\epsilon}  \times \textrm{travel time perturbation extracted from }   C_\epsilon  \right\} .
\end{equation}
The limit $\epsilon \rightarrow 0$ ensures that the above calculation is meaningful irrespective of the level of noise in $C$. For the $s$-th wavepacket,  we have
\begin{equation}
\tau_s    = \lim_{\epsilon \rightarrow 0}   \mathop{\mathrm{argmin}}_{t'}  
\int_{\textrm{win}_s} 
\left[     \epsilon C(t)+ (1-\epsilon) C_{\rm ref}(t) -    C_{\rm ref}(t-\epsilon t')  \right]^2  \id t,
\end{equation}
where the integral is over a time interval $\textrm{win}_s$ centered on the $s$-th wavepacket.
The minimization can be carried out analytically to obtain 
\begin{equation}
\label{eq.ttdef}
\tau_s    = \int W_s(t) \left[    C(t) - C_{\rm ref}(t) \right]  \id t,
\end{equation}
where the function $W_s$ is smooth and depends only on $C_{\rm ref}$ and its first derivative $C'_{\rm ref} = \id C_{\rm ref} / \id t$; see  \citet{Gizon2004} and Section~\ref{sec:methods}. 
This definition of travel time, Equation~(\ref{eq.ttdef}), is extremely robust to noise and is straightforward to compute.
Furthermore, the travel-time perturbation is linear in the perturbation to the autocorrelation function. This property ensures a simple connection between travel-time measurements and perturbations to the stellar model. 
Using travel-time kernels that describe the sensitivity of the measurements to localized changes in the solar interior \citep[e.g.,][]{Gizon2002, Gizon2017, Fournier2018}, we can interpret the seismic data by  solving a linear inverse problem. In addition, this definition of travel time enables us to model the noise covariance matrix \citep{Gizon2004, Fournier2014}. 

The paper is organized as follows. In Section~\ref{sec:methods}, we present the VIRGO/SPM data that cover the last two solar cycles and we describe the data analysis. In Section~\ref{sec:results}, we measure the solar seismic travel times over segments of the data ($T=90$ days) and show that the travel times change with the solar cycle. The implications of our findings  for the detection of activity cycles on other stars are discussed in Section~\ref{sec:discussion}.

\section{Method}
\label{sec:methods} 
\subsection{Observations} \label{sec:data}
\begin{figure*}
\centering   \includegraphics[width=1.0\textwidth]{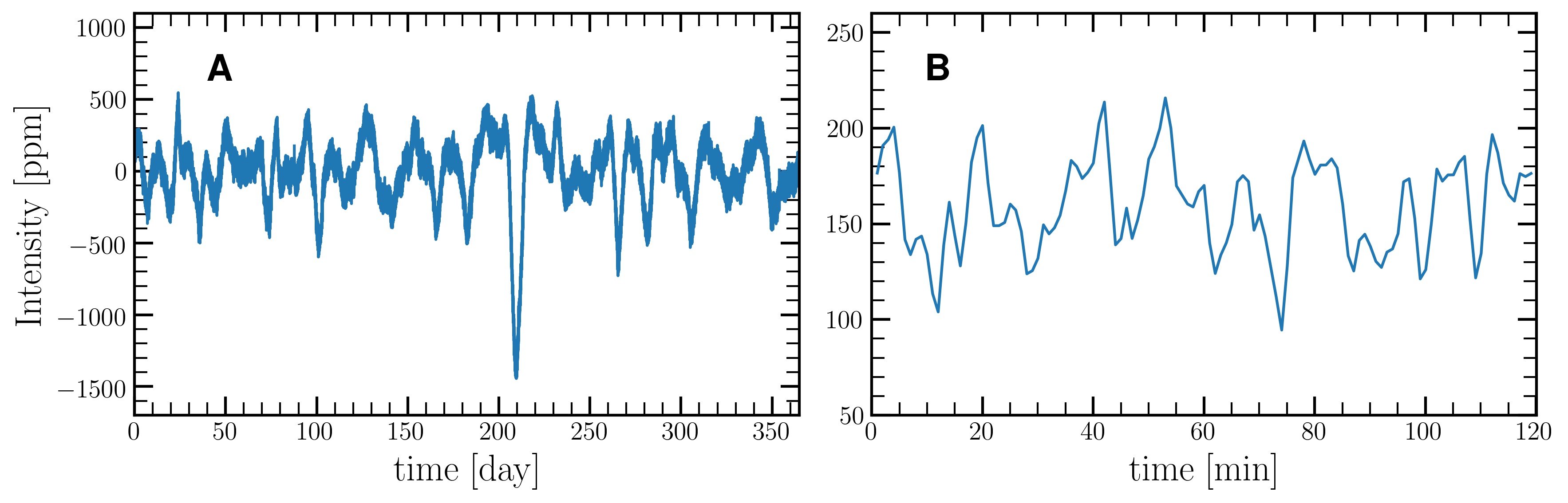}
      \caption{  One-year \textbf{(A)} and   two-hour \textbf{(B)}  segments of  VIRGO/SPM brightness data (red channel). In both plots, the origin of time is 28 March  2014.}
         \label{fig:Figure_1}
   \end{figure*}

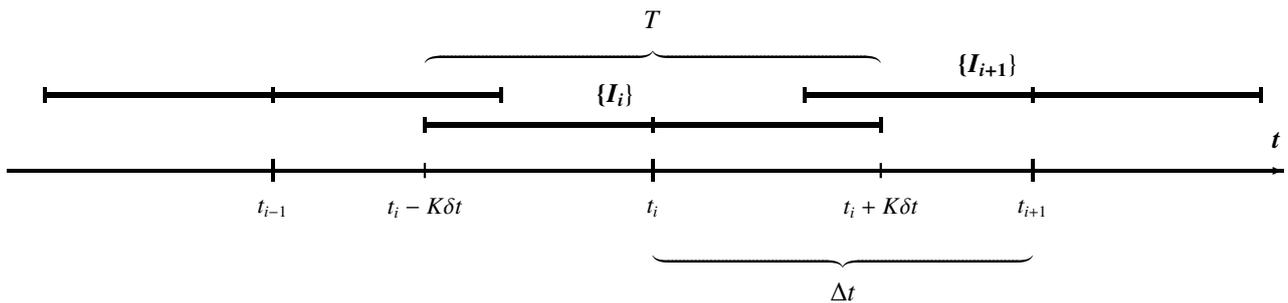
\begin{figure*}
\setlength{\unitlength}{1cm}
\begin{picture}(15, 4)
\put(17.2,2.4){\makebox(0,0){$\boldsymbol{t}$}}
\put(4, 1.5){\makebox(0,0){\small ${t_{i-1}}$}}
\put(9,1.5){\makebox(0,0){\small ${t_{i}}$}}
\put(14,1.5){\makebox(0,0){\small ${t_{i+1}}$}}
\put(6.0,1.5){\makebox(0,0){\small ${t_{i}-K\delta t}$}}
\put(12.0,1.5){\makebox(0,0){\small ${t_{i}+K\delta t}$}}

\put(8.5, 3){\makebox(0,0){$\boldsymbol{\{I_{i}}\}$}}
\put(13.4, 3.4){\makebox(0,0){$\boldsymbol{\{I_{i+1}}\}$}}

\put(6.0,3.5){\makebox(6,0){\downbracefill}}
\put(9.0,4.0){\makebox(0,0){${T}$}}

\put(9, 0.8){\makebox(5,0){\upbracefill}}
\put(11.5, 0.4){\makebox(0,0){${\Delta t}$}}

\linethickness{0.7 mm}
\put(1,3.0){\line(1,0){6}} 
\put(6,2.6){\line(1,0){6}}  
\put(11,3.0){\line(1,0){6}}  

\linethickness{0.4 mm}
\put(4.,2.9){\line(0,1){0.2}} 
\put(9.0,2.5){\line(0,1){0.2}} 
\put(14.0,2.9){\line(0,1){0.2}} 

\put(6.0,2.5){\line(0,1){0.2}} 
\put(12.0,2.5){\line(0,1){0.2}} 
\put(1.0,2.9){\line(0,1){0.2}} 
\put(7.0,2.9){\line(0,1){0.2}} 
\put(17.0,2.9){\line(0,1){0.2}} 
\put(11.0,2.9){\line(0,1){0.2}} 

\linethickness{0.4 mm}
\put(4., 1.85){\line(0,1){0.3}}
\put(9, 1.85){\line(0,1){0.3}}
\put(14, 1.85){\line(0,1){0.3}}
\put(0.5,2){\vector(1,0){16.8}}

\linethickness{0.3 mm}
\put(6.0, 1.9){\line(0,1){0.2}}
\put(12.0, 1.9){\line(0,1){0.2}}

\end{picture}
  \caption{Schematics showing how the data segments $\{I_{i}\}$ are constructed from the original time series of the observed intensity. }
         \label{fig:Figure_2}
 \end{figure*} 
We use Sun-as-a-star observations from the VIRGO experiment on board the Solar and Heliospheric Observatory (SOHO). The VIRGO/SPM instrument is a three-channel full-beam Sun photometer
\citep{Frohlich1995, Frohlich1997}.  It measures solar brightness variations in the continuum  through  three filters that are  centered
 on wavelengths 402~nm (blue), 500~nm (green), and 862~nm (red)  with a 
 temporal cadence of $\delta t = 1$~min. The bandwidth of each filter is 5~nm. 
The response functions of the three channels are located near the base of
the solar photosphere, within $\pm 10$~km of the  $\tau_{500~\mathrm{nm}}=1$  surface \citep[see ][for details]{1998A&A...335..709F}. The blue
and green channels are  more sensitive to the lower part of the solar
atmosphere,  the red channel to higher layers
\citep{Jimenez_2005}. 

The VIRGO/SPM \footnote{The VIRGO/SPM datasets are  available at  \url{http://irfu.cea.fr/dap/Phocea/Vie_des_labos/Ast/ast_visu.php?id_ast=3581}} instrument has been observing the Sun continuously since January 1996, except for   two long gaps.  In the summer of 1998, a gap of about 100 days was due to a rotation  maneuver of the spacecraft that went wrong.  The second gap of about one month occurred in January 1999 as a result of spacecraft problems.  In addition,  5--8\% of the data are not usable depending on the channel \citep[e.g.,][]{Jimenez2002}. The data that  we  use here cover   22 years  from 23  January 1996 to 26 June 2018 (solar cycles 23 and 24). The data are detrended and gaps are filled using linear interpolation. 
Two segments of the data are selected  in Figure~\ref{fig:Figure_1} to display the temporal variations on the rotation timescale  and  on the p-mode oscillation timescale.

\subsection{Data analysis} \label{sec:analysis}
\subsubsection{Autocovariance function}
    \begin{figure*}
   \centering
   \includegraphics[width=1.0\textwidth]{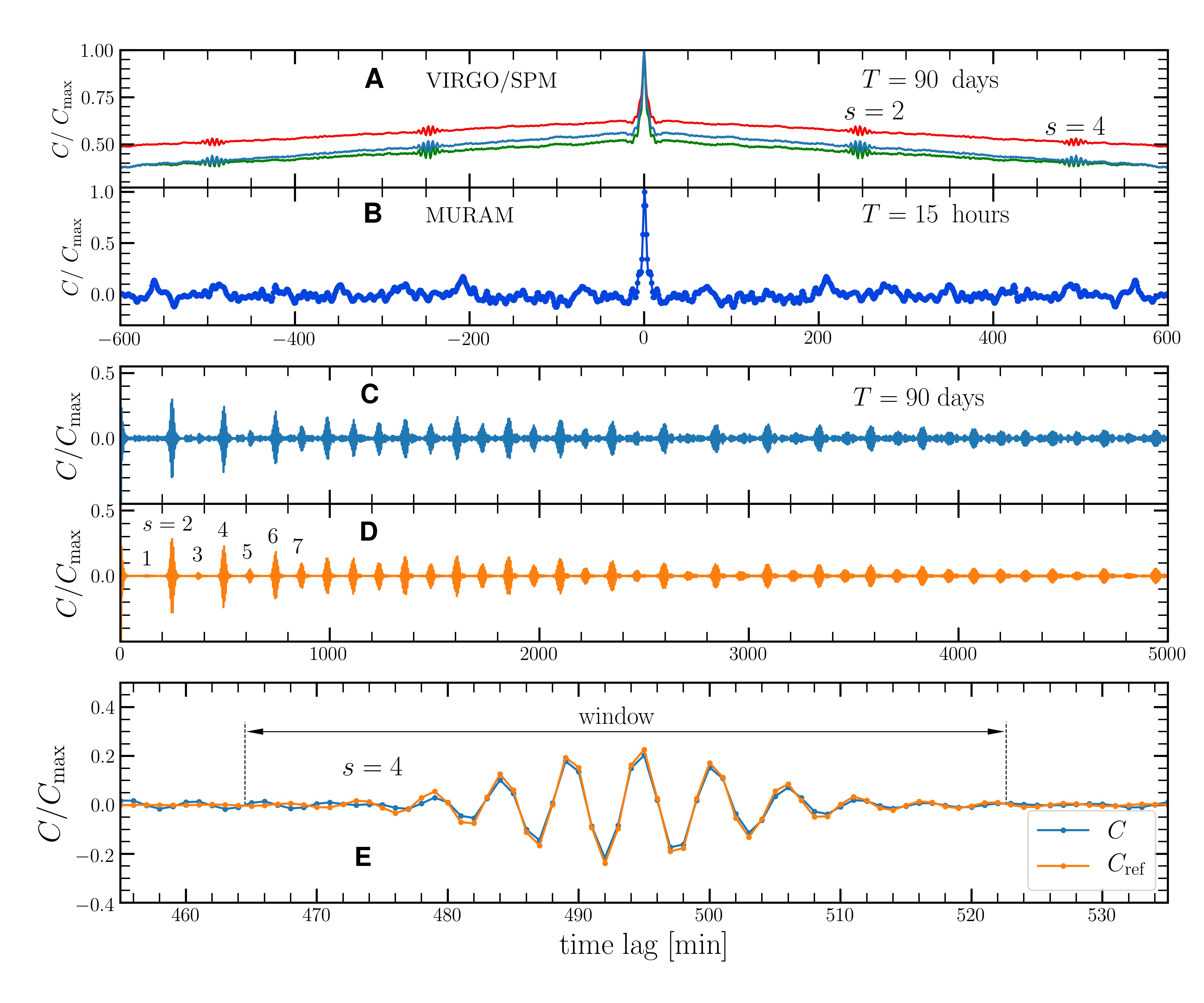}
      \caption{Autocovariance functions of intensity time series.  \textbf{(A)} Autocovariance function of the VIRGO/SPM data ($T=90$~d). The red, green, and blue curves correspond to the three channels. 
       \textbf{(B)} Autocovariance  of MURaM solar convection simulations ($T=15$~hr). 
      \textbf{(C)} Autocovariance function for the p-mode data computed for a segment of VIRGO data with  $T=90$~days (red channel). \textbf{(D)} Reference autocovariance function obtained by averaging over $M=91$ consecutive segments covering 22~years. The first seven wavepackets are labeled  $s=1$ through $7$.  
      The curves in panels (C) and (D) are computed from filtered time series (Gaussian filter centered on 3 mHz with a full width of 2 mHz).
       \textbf{(E)} Zoom  onto wavepacket $s=4$  and  width of the temporal window function $f_4(t)$.
       }
         \label{fig:Figure_3}
   \end{figure*}
Let us consider the intensity, $I(t)$, measured in one of the VIRGO channels at time cadence $\delta t$. We split the time series into $M$ time intervals of equal duration $T=(2K+1) \delta t$, specified through an integer $K$. 
For example, for $K=64800$ we have $T=90$ days. 
The central time of each segment of the data is 
\begin{equation} \label{eq:t_i}
  t_i  = t_{00} +  i\ \Delta t , \quad   0 \le i \le M-1 ,
\end{equation}
where $t_{00}$ is a reference start time and $\Delta t$ is the sampling (e.g., 2 months). These segments may overlap.
Nonoverlapping contiguous segments have $\Delta t = T$.
In each segment  of the data, time is conveniently specified by two indices, $i$ and $k$: 
\begin{equation}
    t_{ik} =  t_i  + k \delta t   
,\end{equation}
with  $0 \le i \le M-1$  and $-K \le k \le K$. Over each time segment, we  define the time series $\{ I_i \}$ such that
\begin{equation}
I_{i}( k \delta t ) :=  I(t_{ik}),  \qquad -K \le k \le K .
\end{equation}
Figure~\ref{fig:Figure_2} summarizes the notations.

To each segment of the data, we apply a Fourier transform: 
\begin{equation}
\hat{I}_{i}(\omega_j) =  \sum_{k=-K}^{K} I_{i}(k\ \delta t) e^{-\mathrm{i} \omega_j k \delta t}, 
\end{equation}
where   $\omega_j = j\Delta \omega$ is the frequency and  $\Delta \omega = 2\pi/T$ is the frequency resolution. The autocovariance function is given by
\begin{equation} 
C_{i}(k \delta t) = \frac{1}{T}\sum_{j=-K}^{K} \left|   I_{i}(\omega_j)   \right|^2 e^{\mathrm{i} \omega_j k \delta t}
,   
\end{equation}
where $k\delta t$ is the correlation time lag. 

Figure~\ref{fig:Figure_3}A shows the autocovariance functions calculated for time series in the red,  green, and blue VIRGO/SPM channels covering the same $T=90$ day  segment in 2018 from 28 March  to  26 June. 
The p modes are clearly seen near time lags $250$~min and $500$~min (the wavepackets with $s=2$ and $s=4$). 
At zero time lag, the autocovariance function has a sharp peak with a width of $\sim 20$ min due to the solar granulation. 
This granulation peak at zero time-lag is well reproduced in  Figure~\ref{fig:Figure_3}B 
using a numerical simulation of solar magnetoconvection simulations in a small Cartesian box \citep{2013A&A...558A..49B}.

{To focus on the p modes, we filter the data by multiplying with a Gaussian filter $F(\omega_j)$ centered at 3~mHz with a full width of 2~mHz:
\begin{equation} 
C_{i}(k\ \delta t) = \frac{1}{T}\sum_{j=-K}^{K} \left|   I_{i}(\omega_j)  F(\omega_j) \right|^2 e^{\mathrm{i} \omega_j k \delta t} ,   
\end{equation}
where $k\delta t$ is the correlation time lag.  We separately filter the positive and negative frequency domains using a Gaussian filter with a full width at half maximum (FWHM) of $2$ mHz centered on about $3$ mHz and $-3$ mHz, respectively. In Figure~\ref{fig:Figure_3}C, we show the autocovariance function computed with the filtered data collected over 90 days. 
   
Next, we construct a reference autocovariance function by averaging $C_i$ over all the time intervals: 
\begin{equation} \label{eqn:c_tau_ref}
C_\mathrm{ref} = \frac{1}{M}\sum_{i=0}^{M-1} C_i.
\end{equation}
In Figure~\ref{fig:Figure_3}D, we show the reference autocovariance function computed by averaging over $M=91$ segments of data covering 22 years  in total. In the time-lag range from $0$ to $3.5$ days, there are about $40$ p-mode wave packets.  To each wave packet, we assign an index $s$ equal to the number of skips that waves forming the given wave packet have before coming back to the original point. We denote the total number of wave packets used in the analysis as $N_\mathrm{skips}$. The first wave packet,  denoted $s=1$, arrives with a time lag of $ 1/ \Delta \nu$, where $\Delta \nu$ is the large frequency separation:
\begin{equation}
\frac{1}{\Delta \nu} = 2 \int_0^{R_\odot} \frac{1}{c}\ {\rm d}r  \approx 123~ \textrm{min}, 
\end{equation}
where $c(r)$ is the sound speed profile and $R_\odot$  the solar radius.  During this time,  waves travel from the surface to their lower turning points and are partially reflected  back to the surface.  Due to the very small amplitude of this wave packet,  we exclude it from the further analysis. The waves that are not reflected back travel further, cross the Sun, reach the surface, get reflected, and travel back.  They arrive with a time lag of $\approx 240$ min,  and we denote them wave packet  $s=2$. With an increase in the number of skips, the penetration depth of p-modes decreases, and depending on the perturbations of the solar structure, p-modes can travel faster or slower. In the autocovariance function, this causes a negative or positive time lag of the wave packet relative to the wave packet with the same skip number in the reference autocovariance. }

\subsubsection{Travel-time measurements} \label{sec:TTVmethod}
{We use the definition for the travel time introduced by
\cite{Gizon2004}. For a given p-mode wave packet $s$, it is a time lag that minimizes the difference between the measured autocovariance $C_i$  and the  reference autocovariance 
$ C_{\rm ref}$:
\begin{equation}
\tau_s(t_i) = \delta t \sum_{k} W_s(k\ \delta t) \left[ C_i(k\ \delta t) - C_\mathrm{ref}(k\ \delta t) \right], 
\end{equation}
where $t_i$ is defined in Equation~\ref{eq:t_i},  and $W_s(k\ \delta t)$ is the weight for the $s$-th wave packet: 
\begin{equation} 
W_s (t) = -\frac{ f_s( t)  \, C^\prime_\mathrm{ref}( t) }{\delta t  \sum_k f_s(k\ \delta t)  [C^\prime_\mathrm{ref}(k\ \delta t)]^2}, 
\end{equation}
where $f_s(t)$ is the window function to isolate  the $s$-th  wave packet, and $C^\prime_\mathrm{ref}$ is the time derivative of the reference autocovariance function (computed in Fourier space). By construction, the window function is  only nonzero in a time-lag interval around the wave packet $s$. In Figure ~\ref{fig:Figure_3}E, we show the measured and the reference autocovariance  for the wave packet $s=4$, and the window function used to isolate it. Following the above approach, we measure travel times for $N_\mathrm{skip}=40$ wave packets with skips from $s=2$ to $41$. }
\begin{figure*}
   \centering
   \includegraphics[width=1\textwidth]{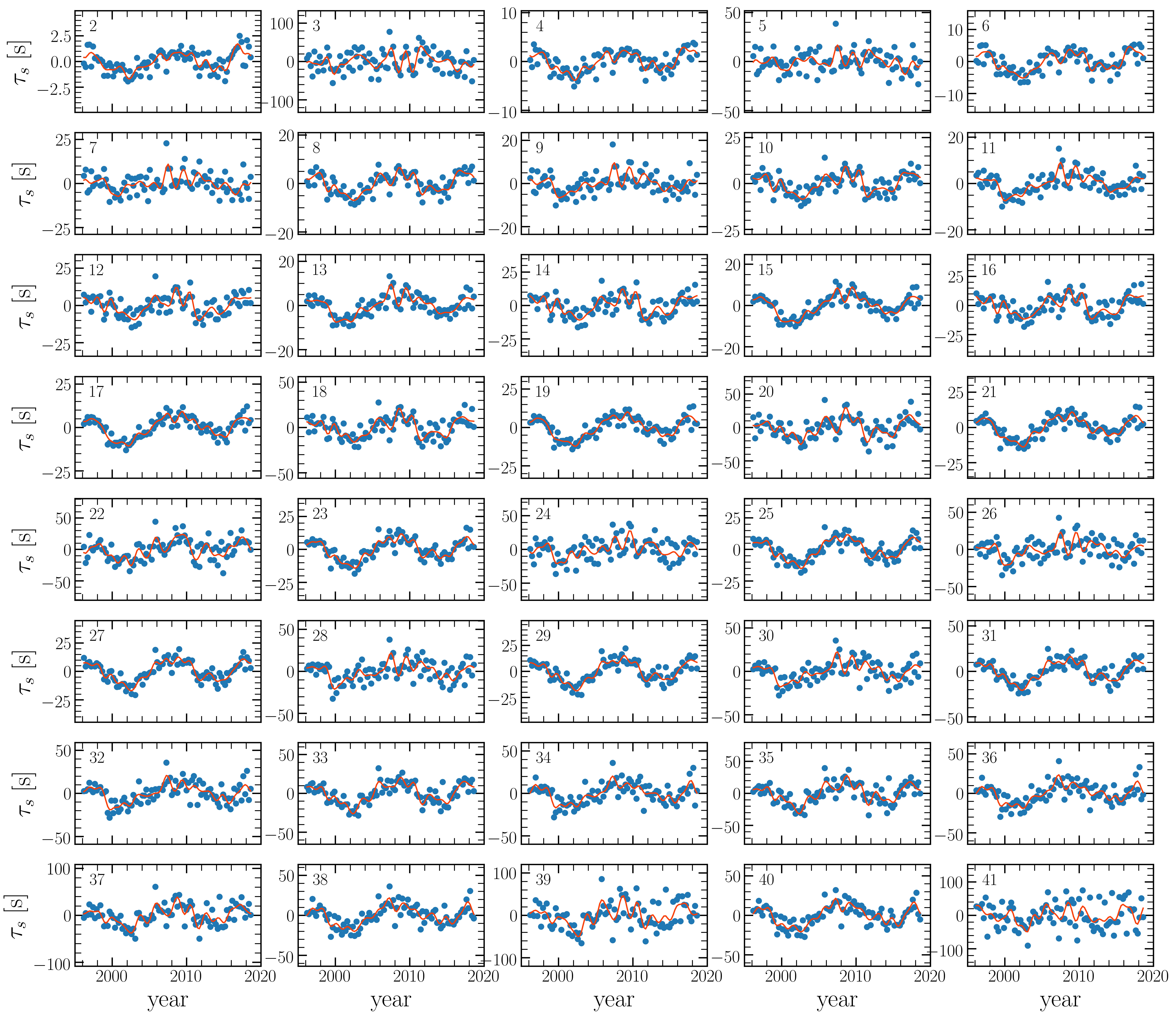}
      \caption{Measured travel times $\tau_s(t_i)$ for all wavepackets with $s \leq 41$. The red curves show the filtered data 
      $\tau_{s,\mathrm{smooth}}(t_i)$ according to Equation~\ref{eq:tau_smooth}. The value of $s$ is written in the top left corner of each plot.  The data analyzed here are from VIRGO/SPM (red channel) divided into non-overlapping segments of $T=\Delta t = 90$ days in  length (see Figure~\ref{fig:Figure_2}).}
         \label{fig:Figure_4}
   \end{figure*}
  \begin{figure*}
\centering
\includegraphics[width=1.0\textwidth]{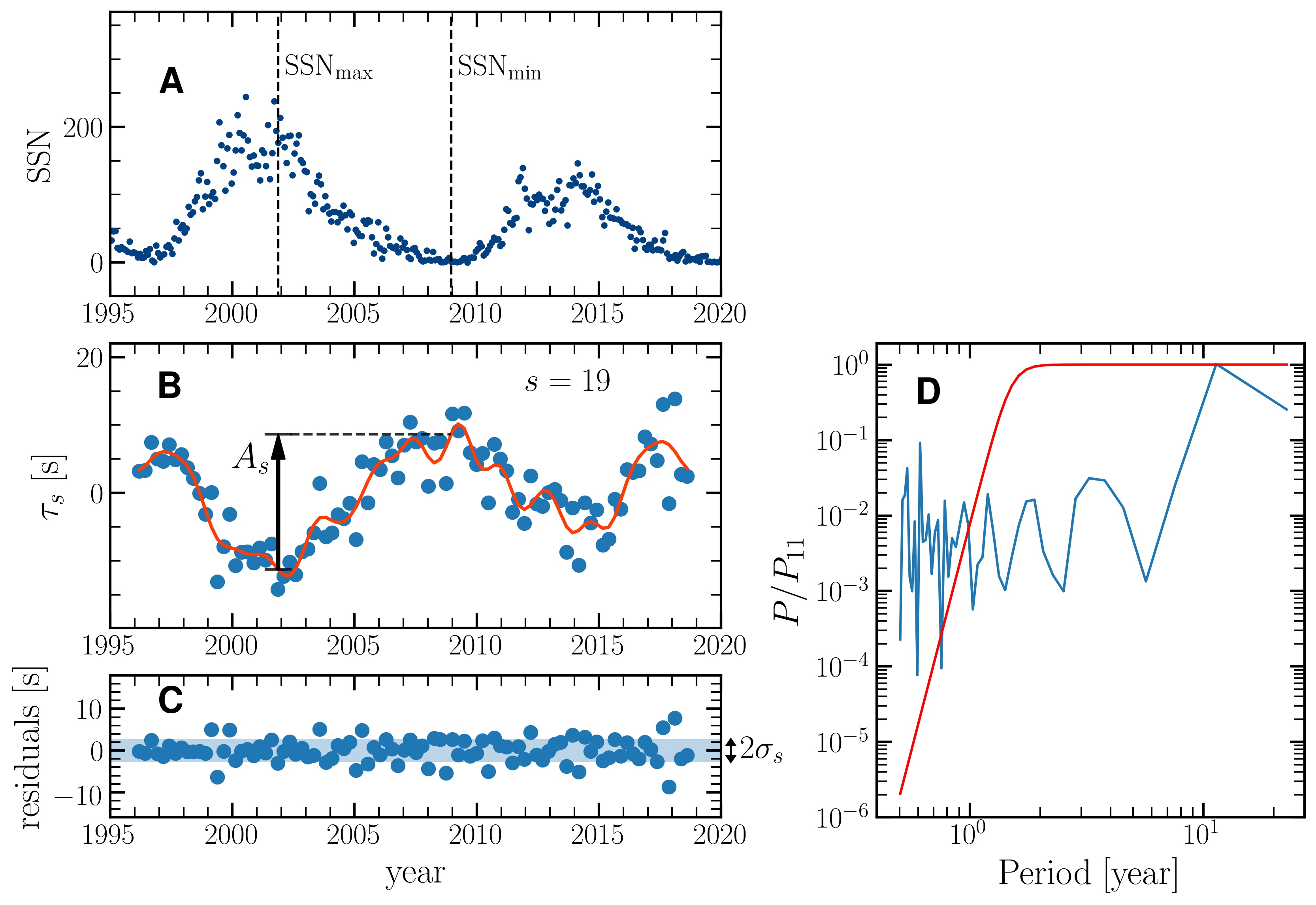}
\caption{ Comparison between the sunspot number (SSN) smoothed over 13 months  \textbf{(A)} and seismic travel times,  $\tau_{s}$, for skip number $s=19$ \textbf{(B)} measured in the red VIRGO/SPM channel data. Vertical dashed lines indicate  cycle maximum and minimum. 
The red line represents the filtered data,  $\tau_{s, \mathrm{smooth}}(t)$, which we use to measure the amplitude of the travel time variations  due to the solar cycle $A_s$ (black arrow). \textbf{(C)} Residuals between $\tau_{s}$ and $\tau_{s,\mathrm{smooth}}$. \textbf{(D)} Power spectrum of $\tau_{s}$, normalized to the power of the  11-year harmonic. A Butterworth filter (red line) is used to extract $\tau_{s, \mathrm{smooth}}(t)$.}
    \label{fig:Figure_5}
\end{figure*}
\section{Results} \label{sec:results}
\subsection{Solar-cycle variations in the travel times}
\label{sec:cycle_in_ttv}
In Figure ~\ref{fig:Figure_4}, we show the travel times $\tau_s$ as functions of $t_i$ for the years 1996--2018 and for each skip number $s\leq 41$. A modulation due to the  11 year solar cycle is seen, except for skips $s=3$ and $5$. Near cycle maximum, the travel times are shorter (negative $\tau_s$), while they are longer near cycle minimum.  The relative variations in travel time induced by the solar cycle depend on the skip number and on the cycle number (23 or 24). For example, the imprint of the sunspot cycle is very easy to see in the travel time variations for the odd skips from 15 to 35. On the contrary, the cycle is difficult to see in the variations of skips 3, 5, 7, 9, 24, 26, 28, and 41. 
The noise in the travel times  clearly depends on the skip number. For example, skips 3, 39, and 41 are particularly noisy.
   
We now wish to measure the amplitude of the  variations due to the solar cycle and its uncertainty in each of the time series $\tau_s$ shown in Figure~\ref{fig:Figure_4}.
For each skip number $s$, we estimate the signal by applying a low-pass filter to the data, such that only periods longer than $1.5$ years remain. This threshold also lets  through  the quasi-biennial variations  \citep[e.g.,][]{Bazilevskaya2014}. 
To filter the data, we use a sixth-order Butterworth filter $H_{6}(\omega, \omega_\mathrm{c}) = (1+(\omega/\omega_{\rm c})^{12})^{-1}$
with a cut-off frequency of $\omega_{\rm c} = 2 \pi/(1.5 \, \mathrm{yr})$:
\begin{equation}  \label{eq:tau_smooth}
\tau_{s,\mathrm{smooth}}(t_i) = \frac{1}{M \Delta t} \sum_{q=0}^{M-1} H_{6}(\omega_q, \omega_\mathrm{c}) \hat{\tau}_s(\omega_q) e^{\mathrm{i} \omega_q t_i},
\end{equation}
where  $\omega_q=2\pi q/(M \Delta t)$ and $\hat{\tau}_s(\omega_q)$ is the Fourier transform of the travel times.  For Figure~\ref{fig:Figure_4}, we have $M=91$ and $\Delta t=90$~days.
The noise in the travel times can be estimated by computing the standard deviation of the residuals:
\begin{equation} \label{eq:noise_tau_s}
\sigma_s = \sqrt{\frac{1}{M-1} \sum_{i=0}^{M-1} \Big[n_s(t_i)\Big]^2}
,\end{equation}
with
\begin{equation} \label{eq:sigma}
n_s(t_i) = \tau_s(t_i) - \tau_{s,\mathrm{smooth}}(t_i) .
\end{equation}
We define the times $\tmax$ and  $\tmin$ 
to correspond to  the solar cycle maximum in November 2001 and the cycle minimum in December 2008. 
Next, we define the travel-time difference: \begin{equation}
A_s =  - \tau_{s, \mathrm{smooth}} (\tmax) + \tau_{s,\mathrm{smooth}}(\tmin). 
\label{eq:amplitude}
\end{equation}
By construction,  we expect  $A_s > 0$ when the variations in travel times are dominated by the solar cycle perturbations. For example, in Figures~\ref{fig:Figure_5}A and B  the  variations of $\tau_{19}(t_i)$ are very clearly anti-correlated with the sunspot number. Figure~\ref{fig:Figure_5}C and D show the residual high-frequency noise in the travel times and how the travel times are filtered for estimation of $A_{19}$.  

\begin{figure*}
   \centering
   \includegraphics[width=\textwidth]{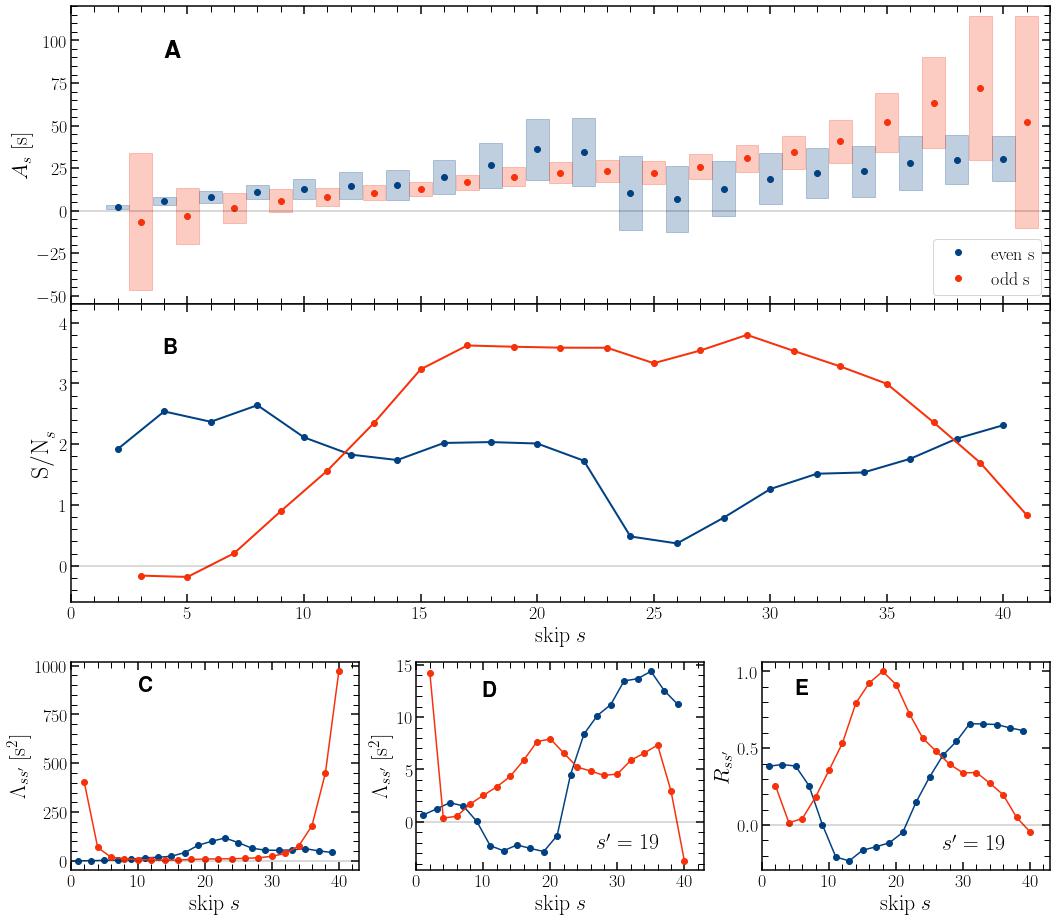}
      \caption{
      Detectability of the solar cycle in individual skip travel times. 
      \textbf{(A)} 
      Amplitudes of travel time variations due to the solar cycle, $A_s$,  versus skip number $s$ (see Equation~\ref{eq:amplitude}), with error bars of $\pm \sqrt{2} \sigma_s$ (see Equation~\ref{eq:var_individual_skips}).
      \textbf{(B)} 
      Signal-to-noise ratios $\mathrm{S/N}_s=A_s/(\sqrt{2} \sigma_s)$.   \textbf{(C)} Variance of travel-time noise  $\sigma_s^2 = \Lambda_{ss}$. \textbf{(D)} Plot of $\Lambda_{s, s'=19}$  and \textbf{(E)} plot of   $R_{s,s'=19}$. In all panels, the blue and red points show the even and odd values of $s$.}
         \label{fig: Figure_6}
   \end{figure*} 

According to the definition of $A_s$, its variance is given by
\begin{equation}
 \mathrm{Var}({A_s}) = 2 \sigma_s^2, 
 \label{eq:var_individual_skips}
\end{equation} 
where $\sigma_s$ is given by Equation~\ref{eq:noise_tau_s}.
Then, for each skip $s$, we define the ratio between the signal due to the solar cycle and the noise level as
\begin{equation}
    \mathrm{S/N}_s = A_s / (\sqrt{2} \sigma_s).
  \label{eq:SNR_sun_individual_skips}
\end{equation}
In Figure~\ref{fig: Figure_6}A and B, we show $A_s$, $\sqrt{2} \sigma_s$,  and $\mathrm{S/N}_s$ for skip numbers $s=2$ to $41$. We find that they strongly depend on skip number $s$. Surprisingly,  the highest signal-to-noise ratios ($\mathrm{S/N}_s > 3$) are for odd skip numbers from $s=15$ to $35$.  Skip numbers $s=3$, $5$, $7$, $24$ and $26$ are very noisy ($\mathrm{S/N}_s < 0.5$).

\subsection{Noise correlations in travel times} \label{sec:noise_correlations}
The noise covariance matrix  for $\tau_s$ is given by 
\begin{equation}
\Lambda_{ss'} = \mathbb{E}[ n_s n_{s'} ] \simeq \frac{1}{M-1} \sum_{i=0}^{M-1} n_s(t_i) n_{s'}(t_i), 
\end{equation}
where $s$ and $s'$ belong to  $\{2, 3, \cdots N_\mathrm{skip}+1\}$.
The noise correlation matrix is then $R_{ss'} =  \Lambda_{ss'}/{\sigma_s \sigma_{s'}}$.
Figures ~\ref{fig: Figure_6}C-E  show cuts through the matrices $\Lambda$ and $R$.  As mentioned in the previous section, the variance of the noise depends strongly on the value and parity (odd or even) of $s$  (Figure~\ref{fig: Figure_6}C). 
At a fixed $s'$, the noise covariance $\Lambda_{ss'}$ and the noise correlation  $R_{ss'}$  depend strongly on the skip number $s$. For example, for $s'=19$, the correlation is large and positive  ($R_{s, 19}>0.6$) for odd skips from $s=15$ to $23$ and for even skips from $s=32$ to $40$, but the correlation is negative for even skips from $s=12$ to $22$   (Figure~\ref{fig: Figure_6}D and E).

\begin{figure*}
\centering
\includegraphics[width=1.0\textwidth]{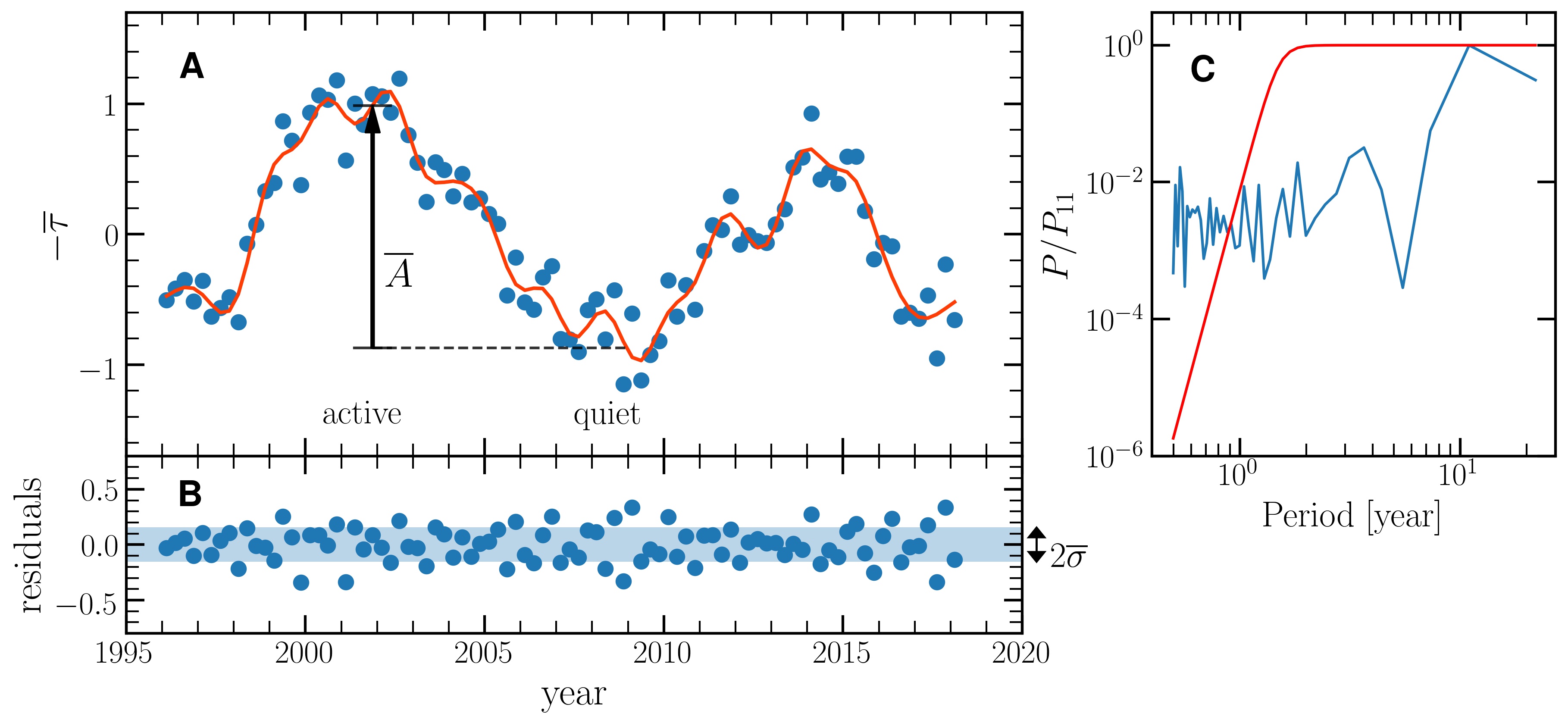}
\caption{ The average seismic travel time. \textbf{(A)} Measured average travel time $\overline{\tau}$ (blue points) using $N_\mathrm{skip} = 40$ 
$\overline{\tau}_{\mathrm{smooth}}$ and the filtered data (red line). 
The vertical arrow defines the amplitude of the 11 year  cycle measured between active and quiet years. 
The VIRGO data were divided into segments of length $T=90$ days with a sampling time of $90$ days. 
\textbf{(B)} Residuals between $\overline{\tau}$ and $\overline{\tau}_{\mathrm{smooth}}$. \textbf{(C)} Power spectrum of $\overline{\tau}$ (blue  curve) normalised to $P_{11} = P$(period = 11 yr). The red curve shows the  Butterworth filter with  a cutoff at $1.5$~yr. 
}
    \label{fig:Figure_7}
\end{figure*}
\begin{figure*}
\includegraphics[width=1.0\textwidth]{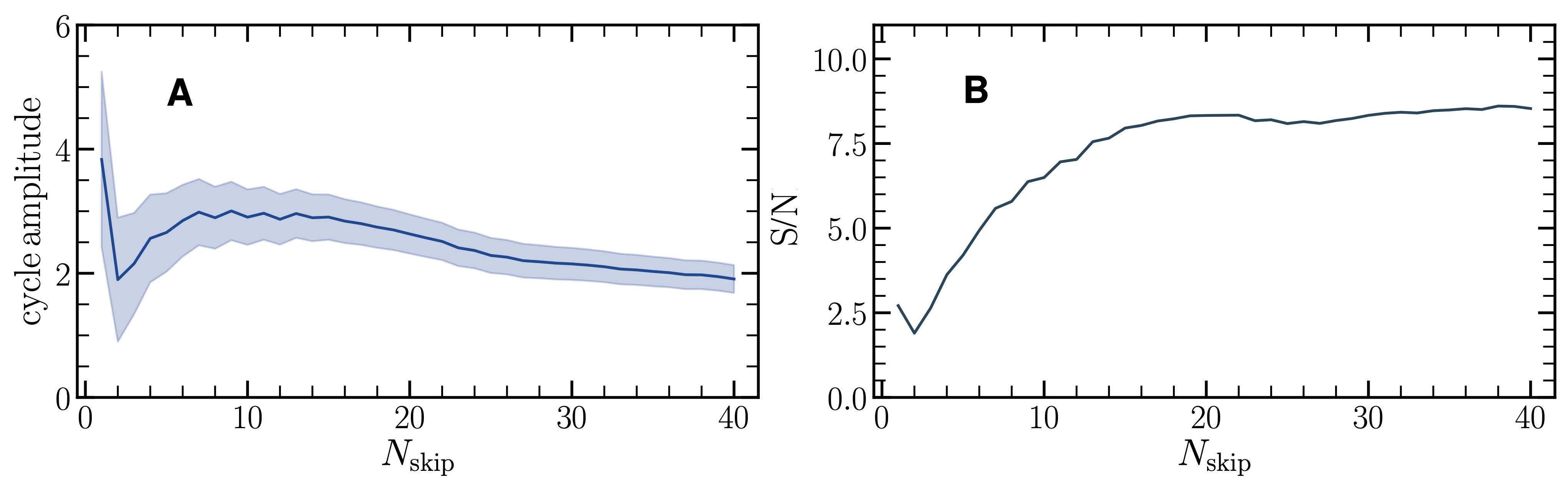}
\caption{Detectability of the solar  cycle in the average VIRGO travel times. \textbf{(A)} Cycle amplitude $\overline{A}$  versus  number of skips $N_{\rm skip}$ used in the average. The colored area outlines the uncertainty  $\pm \sqrt{2} \overline{\sigma}_s$ (Equation~\ref{eq:var}). \textbf{(B)} S/N  of the 11 year cycle as a function of  $N_\mathrm{skip}$. The data analyzed here were divided into segments of $T=90$ days in  length with a sampling time of $90$ days.}
    \label{fig:Figure_8}
\end{figure*}
 \begin{figure}
    \includegraphics[width=0.5\textwidth]{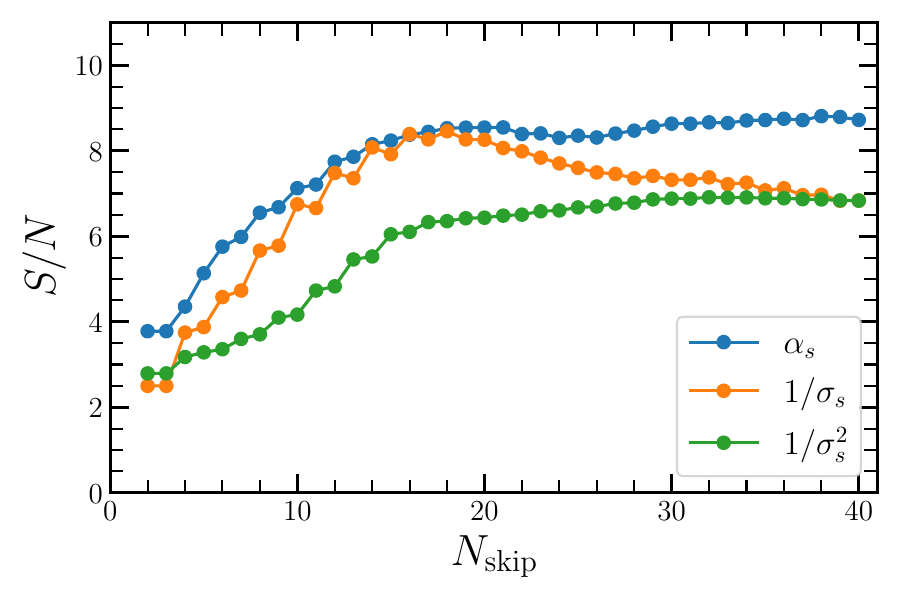}
    \caption{Signal-to-noise ratio associated with different averages of the  seismic travel times versus maximum skip number $N_{\rm skip}$ used in the average. For all $N_{\rm skip}$, the weights $\alpha_s$ used in the average (blue) provide a S/N that is higher than the weights proportional to $1/\sigma_s$ (orange) and $1/\sigma^2_s$ (green).
    } \label{referee:comparisson_snrs}
\end{figure}
  \begin{figure*}
\includegraphics[width=1.0\textwidth]{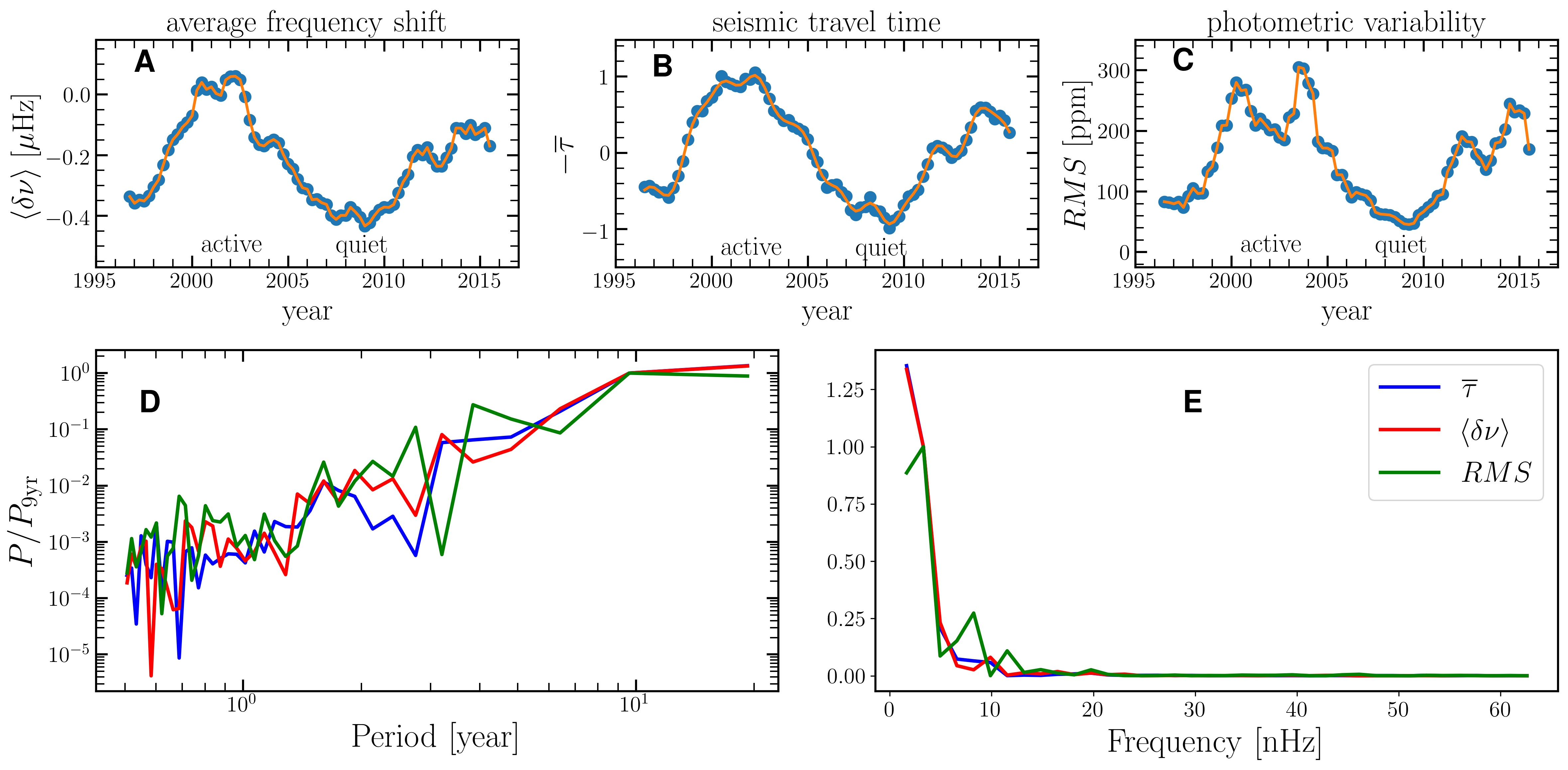}
\caption{Solar cycle variations in the seismic average mode frequency shift from  \cite{Howe2017} \textbf{(A)}, the seismic average travel time \textbf{(B)}  and 
   photometric  variability \textbf{(C)}. The orange curve connects the data points. The data analyzed here was divided into  overlapping time segments of $T=1$~yr in length and mid-points separated by  $\Delta t = 91.25$~d. \textbf{(D-E)} Power spectra of the activity 
observables shown in panels A-C. All power spectra are normalized to the power at  frequency $1/(9\ \textrm{yr})$. The abscissa shows the period in the left panel and the frequency in the right panel.}
    \label{fig:Figure_9}
\end{figure*}

\subsection{Travel-time averages and signal-to-noise ratio} \label{sec:signal-to-noise_ttv_sun}
Here, we construct a linear combination of the travel times that takes the noise correlations into account in order to reduce the overall noise and detect the solar activity cycle with the highest S/N. Minimizing the variance of the noise can be achieved (1) by transforming the travel times to statistical orthogonality and (2) by  applying a straight average \citep[see, e.g.,][]{Kessy2018}. 
To construct a new set of statistically independent travel-time  measurements, we apply the standard transformation
\begin{equation} 
\check{\tau}_s(t_i) = \sum_{s'=2}^{N_\mathrm{skip}+1} \Big( \Lambda ^{-1/2} \Big)_{ss'} \tau_{s'}(t_i),  \label{eq:whitening}
\end{equation}
where $\Lambda_{ss'}^{-1/2}$ is the whitening matrix. By construction, the covariance matrix of the whitened data is the identity matrix:
\begin{equation}
\frac{1}{M-1} \sum_{i=0}^{M-1} \check{\tau}_s(t_i) \check{\tau}_{s'}(t_i) = \delta_{ss'}.
\end{equation}
Next, we compute a straight average of the whitened time series over skips to obtain the average travel time:
\begin{equation}
\meantau =  \frac{1}{N_\mathrm{skip}} \sum_{s=2}^{N_\mathrm{skip}+1} \check{\tau}_s(t_i).
\label{eq:straight_avrage_mean_tau}
\end{equation}
Equations~\ref{eq:whitening} and \ref{eq:straight_avrage_mean_tau} can be combined into one formula:
\begin{equation}
\meantau =  \sum_{s=2}^{N_\mathrm{skip}+1} \alpha_s \tau_s(t_i) \quad \mathrm{with }\  \alpha_s = \frac{1}{N_\mathrm{skip}} \sum_{s'=2}^{N_\mathrm{skip}+1} \Big( \Lambda ^{-1/2} \Big)_{ss'}.
\label{eq:tau_mean}
\end{equation}

The average signal and noise components are 
\begin{eqnarray} \label{eq:mean_model_averaged_ttv}
\overline{\tau}_{\mathrm{smooth}}(t_i) &=& \sum_{s=2}^{N_\mathrm{skip}+1} \alpha_s \, \tau_{\mathrm{smooth}, s}(t_i), 
\\
\overline{n} (t_i) &=& \overline{\tau}(t_i) - \overline{\tau}_{\mathrm{smooth}}(t_i) .
\end{eqnarray}
The standard deviation of the average noise is
\begin{equation}
\overline{\sigma} = \sqrt{\frac{1}{M-1} \sum_{i=0}^{M-1} \Big[\overline{n} (t_i)\Big]^2}.
\end{equation}

Given these average quantities, the solar activity cycle amplitude becomes 
\begin{equation} \label{eq:expectation_A}
 \overline{A} = -\overline{\tau}_{\mathrm{smooth}} ( \tmax)  +\overline{\tau}_{\mathrm{smooth}}( \tmin) ,
\end{equation}
with variance
\begin{equation}
 \mathrm{Var}(\overline{A}) =  2 \overline{\sigma}^2
 \label{eq:var}
\end{equation} 
and 
\begin{equation}
    \mathrm{S/N} = \overline{A} / \sqrt{ \mathrm{Var}(\overline{A}) }.
  \label{eq:SNR_sun}
\end{equation}
Figure~\ref{fig:Figure_7} shows the measured averaged travel time $\overline{\tau}(t_i)$,  the average signal $\overline{\tau}_{\mathrm{smooth}}(t_i)$,  and the noise level $\overline{n} (t_i)$, all obtained after averaging over   $N_\mathrm{skip}=40$ skips. 
The improvement in  the  S/N as a function of skip number is plotted in Figure \ref{fig:Figure_8}. We find that the S/N increases steadily with $N_{\rm skip}$ until  $N_{\rm skip}\approx 20$, and then reaches a plateau. The maximum S/N is approximately 8.

The weights $\alpha_s$ lead to an average travel time, which has the highest S/N. If, instead, we were to chose weights proportional to $1/\sigma_s$ or $1/\sigma_s^2$, the S/N of the average travel time would be significantly lower, as illustrated in Fig.~\ref{referee:comparisson_snrs}. 

\section{Discussion} \label{sec:discussion} 
\subsection{Comparison of seismic travel times with other activity proxies}\label{sect:SNR_seismology_photometry_and_delta_nu}
We use the frequency shifts of $60$ low-degree modes with $l\leq 3$ that were measured by \cite{Howe2017, Howe2018} from the  Birmingham Solar Oscillations Network (BiSON) data between 23 January 1996 and 31 December 2016.  The individual mode frequency shifts were measured for overlapping segments of $T=365$ days in  length with a sampling time of $\Delta t =  91.25$ days (only one-quarter of the data points are statistically independent). The average frequency shifts, weighted by mode inertia, are denoted by $\langle \delta \nu \rangle$ and are plotted in Figure~\ref{fig:Figure_9}E.
In order to directly compare the BiSON frequency shifts to the seismic travel times and the photometric variability, we divided the VIRGO data into the same overlapping segments ($T=365$ days and $\Delta t=91.25$ days). From these data segments, we obtain the average travel time $\overline{\tau}$ with $N_\mathrm{skip}=40$ and the photometric $RMS$. 
\begin{figure*}
\centering
\includegraphics[width=0.9\textwidth]{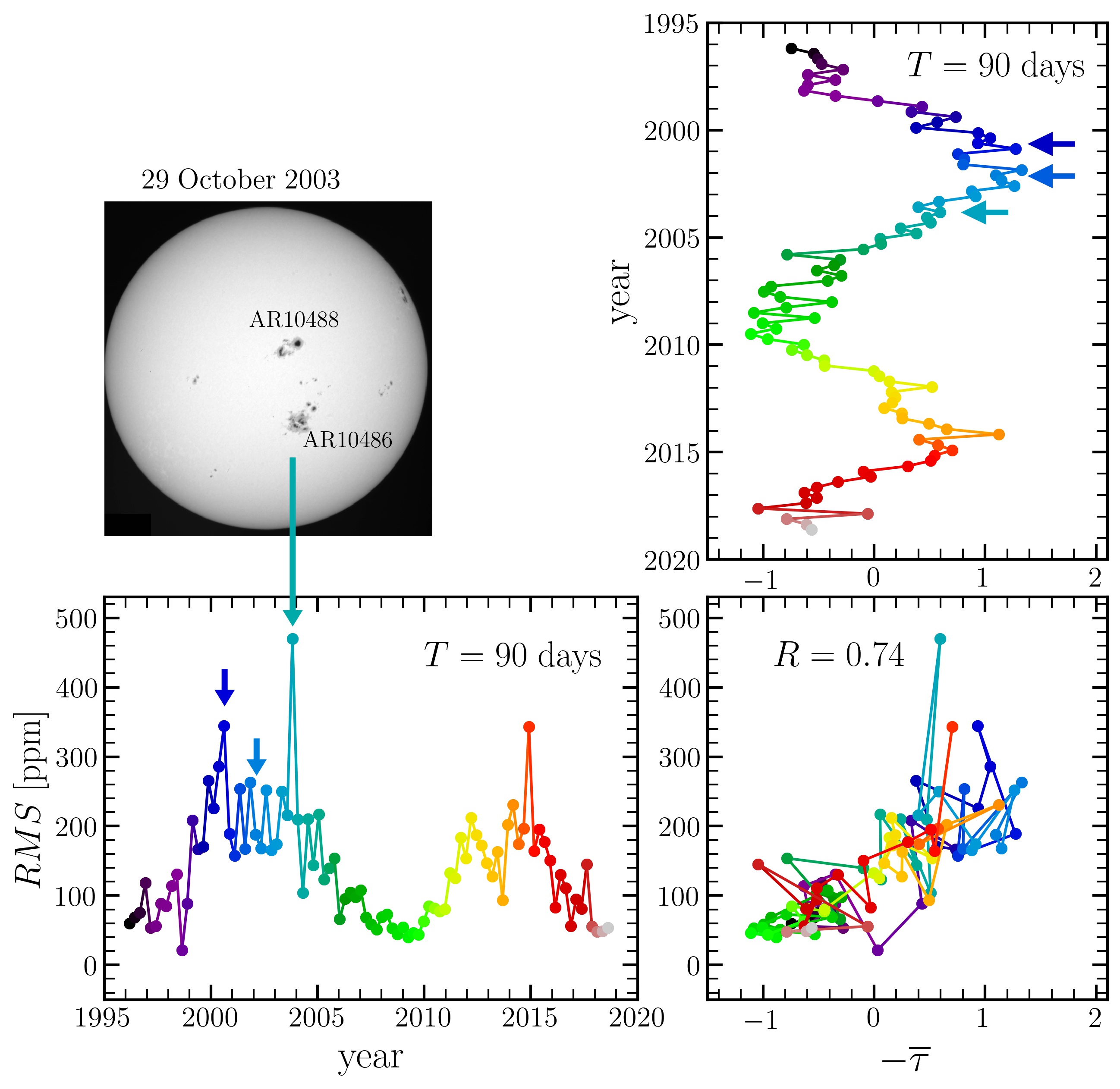}
    \caption{
     Comparison between photometric variability  and  travel-time perturbations.
    \textit{Bottom left panel:} 
    Photometric variability  computed from nonoverlapping time segments of $T=90$~days in duration. The three arrows correspond to three particular times in July 2000, September 2001, and October 2003.
    The strong spike in the data in October 2003 is due to the transit of two large active regions (\textit{top left panel}).
    \textit{Top right panel:} Seismic travel times computed for the same 90 day time segments.
    As above, the three arrows point to July 2000, September 2001, and October 2003.
    \textit{Bottom right panel:} Scatter plot between photometric variability  and seismic travel times. 
    The correlation coefficient between these datasets is $R=0.74$. }
    \label{fig:Figure_10}
\end{figure*}
\begin{figure*}
\centering
\includegraphics[width=0.9\textwidth]{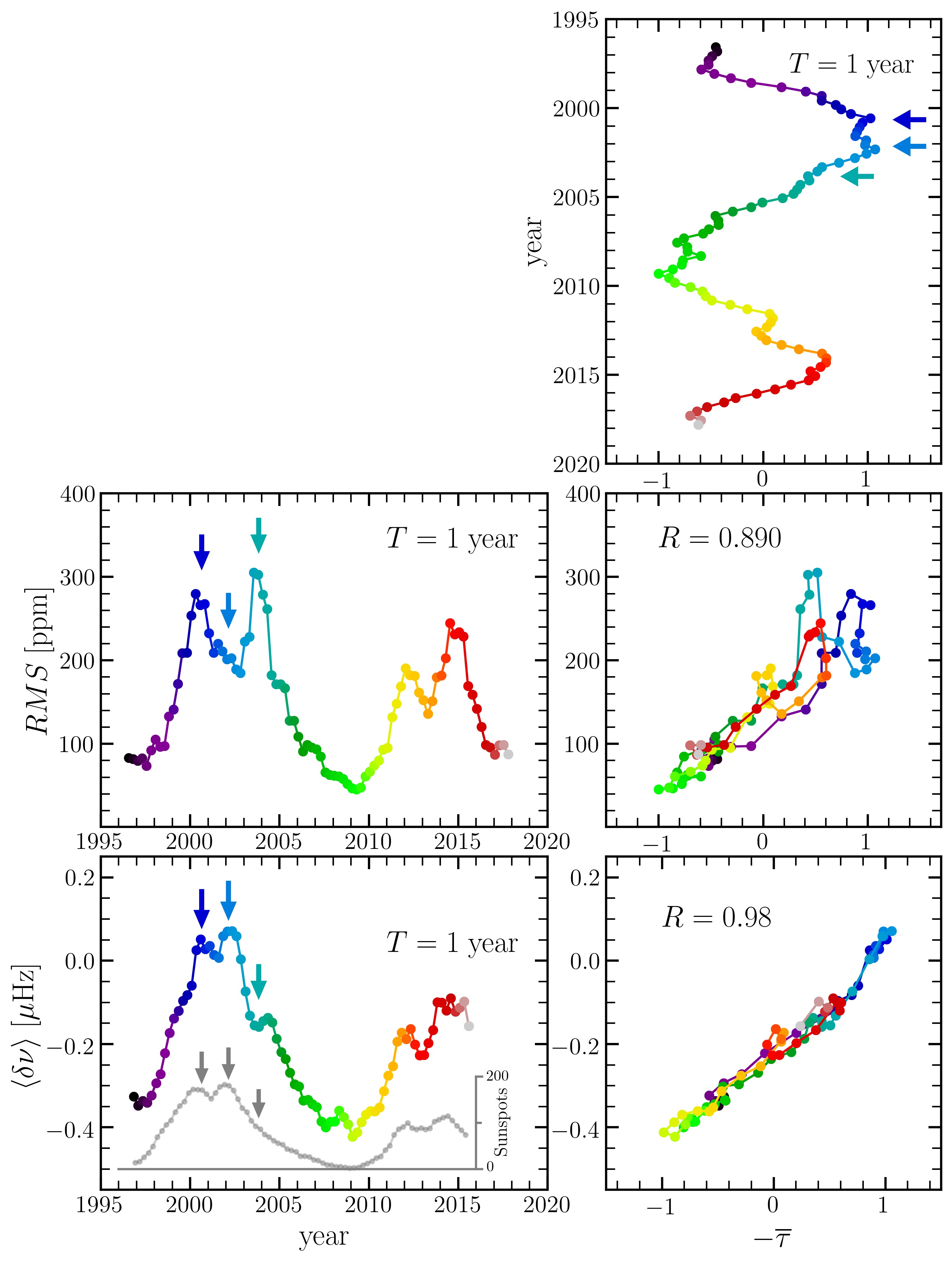}
    \caption{ 
    Comparison between photometric variability, p-mode frequency shifts, and seismic travel times.  The data analyzed here were divided into overlapping segments of $T=365$ days in length with a sampling time of $91.25$ days.
    The three arrows correspond to July 2000, September 2001, and October 2003, as in Figure~\ref{fig:Figure_10}. 
    The bottom right panel shows the scatter plot between travel times and mode frequencies.
    The center-right panel  shows the scatter plot between travel times and photometric variability. In the left bottom panel, the gray line shows smoothed sunspot number.}
    \label{fig:Figure_11}
\end{figure*}
\begin{figure*}
\centering
\includegraphics[width=1.0\textwidth]{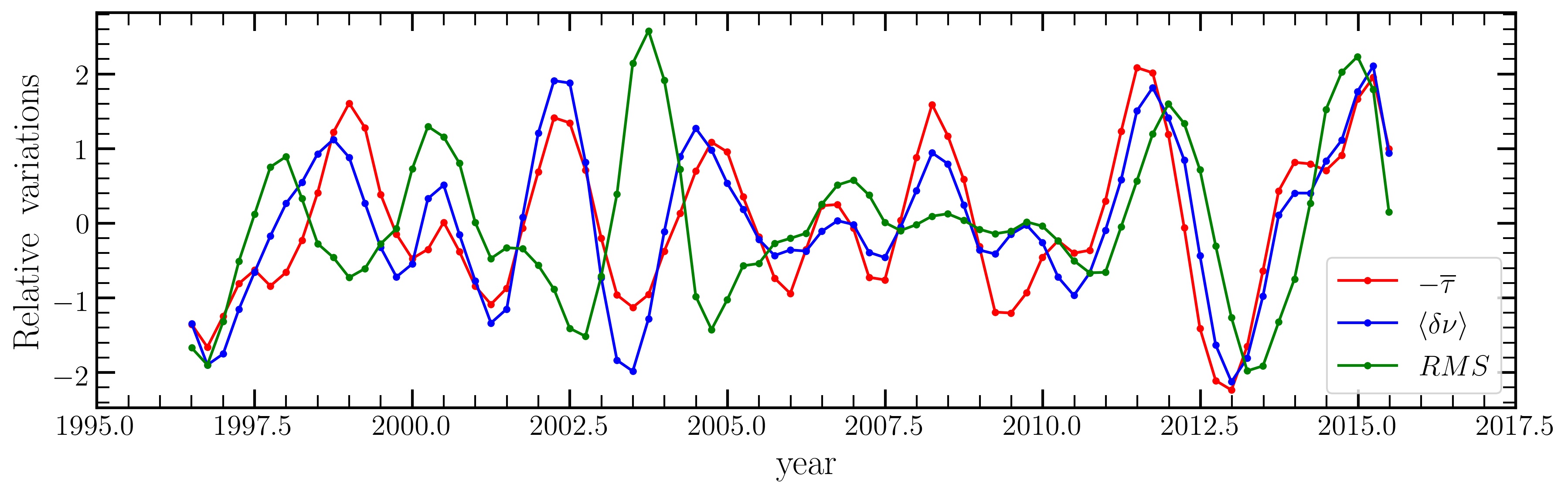}
    \caption{Quasi-biennial variations in $\overline{\tau}$ and $\langle \delta \nu \rangle$, and $RMS$ time series during solar cycles 23 and 24, normalized by their standard deviations. The data analyzed here were divided into overlapping segments of $T= 365$ days in  length with a sampling time of $91.25$ days. The three data sets were filtered in the range $1.5$--$3.5$~yr using a Gaussian filter.  }
    \label{fig:Figure_12}
\end{figure*}
\begin{figure*}
\includegraphics[width=1.0\textwidth]{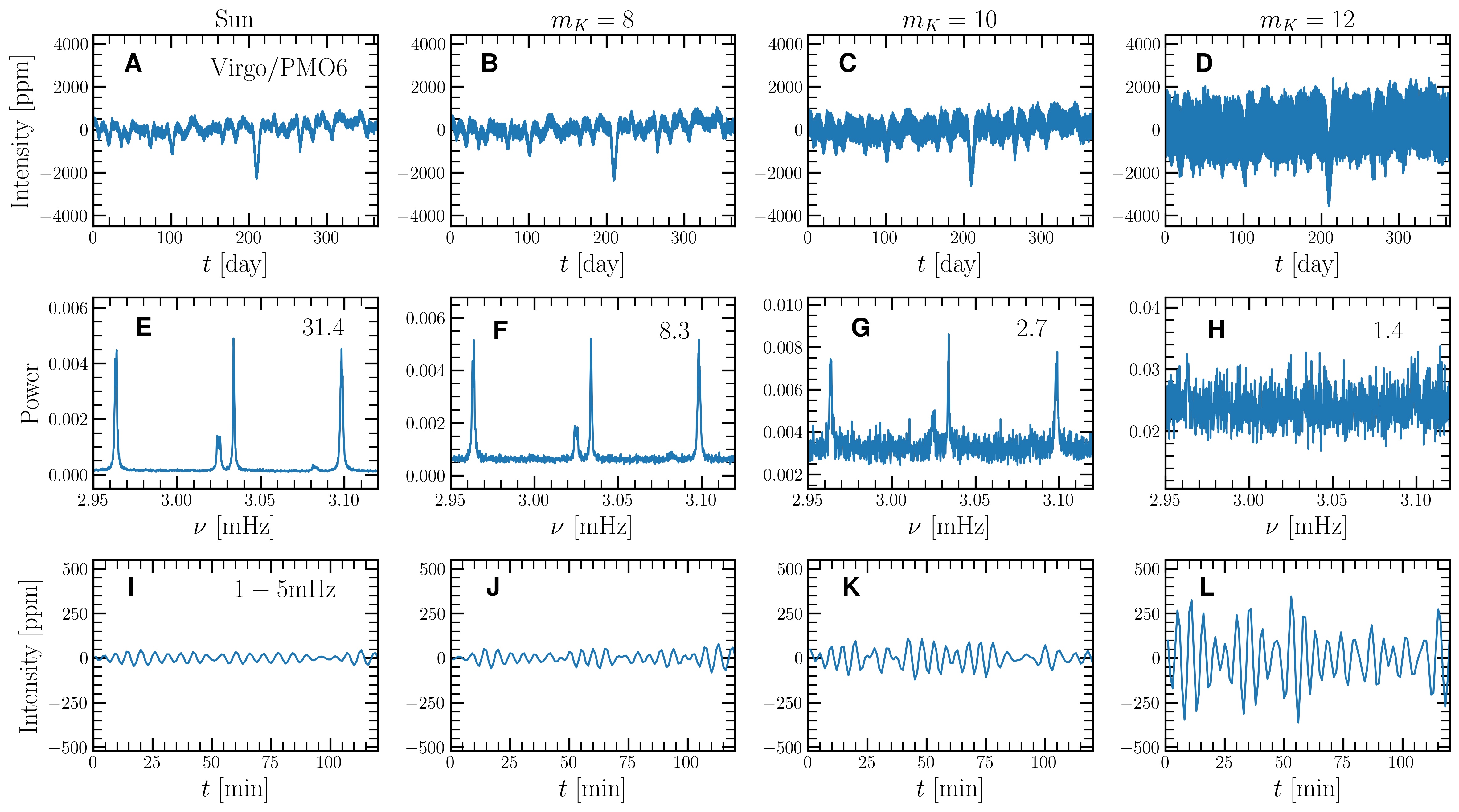}
    \caption{Comparison between the solar VIRGO/SPM observations and simulated Sun-as-a-Kepler-Star data.  Example one-year time series of \textbf{(A)} VIRGO/SPM observations and \textbf{(B--D)} simulated Sun-as-a-Kepler-star data for stars of different visual magnitudes.  \textbf{(E-H)} P-mode power spectra near 3~mHz. The values in the upper-right corners give the power ratios between the height of the ($l=0$) peak at $3.097$~mHz  and the noise background. 
\textbf{(I--L)} Filtered data in the p-mode frequency range 1--5~mHz.
}
    \label{fig:Figure_13}
\end{figure*}

Figure \ref{fig:Figure_9}A-C displays all three data sets:  the average frequency shifts, the average travel times, and the photometric RMS values. 
Variations over the 11 year solar cycle are clearly seen in all three datasets. 
The variations of the average frequency shifts and travel times are more closely related to each other than each one of them is to the $RMS$ data. This last point is also seen when comparing the three power spectra (Figure~\ref{fig:Figure_9}D-E). 
In particular, the $RMS$ data have significant power near periods of $3.8$~yr and $2.7$~yr (with a gap near $3.2$~yr), while the seismic data have a smoother distribution of power for periods above $3.2$~yr.
The noise level for periods below $\sim 2$~years is similar in all three data sets.

Figure \ref{fig:Figure_10} shows the photometric variability and seismic travel times of the Sun using nonoverlapping segments of $T=90$ days in length.
We notice that the  photometric variability time series shows pronounced spikes in July 2000 and October 2003 during Cycle 23, which can be attributed to the transit of large active regions. At the end of the month of October 2003, the increase in photometric variability was particularly notable, reaching almost $500$~ppm due to the combined contributions of two large sunspot groups (NOAA~10488 and NOAA~10486) that shared the same longitude in the rotating frame (see Figure~\ref{fig:Figure_10}, the top left panel). 
This is consistent with the simulations of \citet{Isik_2020}, which show that longitudinal nests of active regions may amplify the brightness variations on other Sun-like stars compared to the Sun.
Moreover, \cite{Pojoga_2002}  reported that a significant fraction of all solar active regions belong to longitudinal clusters on the surface, providing a possible explanation for the larger photometric variability observed during solar maximum than the seismic variability (Figure~\ref{fig:Figure_10}).
Indeed, unlike photometric variability, the seismic waves are not sensitive to the longitudinal distribution of magnetic activity but sense a longitudinal average.

Figure~\ref{fig:Figure_11} shows the correspondence between the photometric variability, mode frequency shifts, and seismic travel times. To reduce random noise, we used overlapping time segments of $T=1$ year in length.  As anticipated, the two seismic measurements are highly correlated ($R=0.98$), although they are not identical. 
We marked three dates on the plot, which correspond to the same active phase of Cycle 23 as in Figure~\ref{fig:Figure_10}. 
Two of these dates correspond to local maxima in the photometric variability, and the middle date corresponds to a maximum for the seismic data (both frequency shift and travel time). The photometric variability exhibits a distinctive double-peak pattern in both Cycles 23 and 24, with the two peaks separated by approximately $3$ years. The seismic data also display local maxima, but the variations are smoother and considerably different from the photometric variability observed during Cycle 23. 
In particular, the seismic data show a peak in September 2001, which has no counterpart in the photometric variability.  We also note that during the rising and  declining  phases of Cycle 23, the relationship between photometric variability and travel times shows hysteresis. 

As demonstrated in the lower-left panel of Figure~\ref{fig:Figure_11}, the p-mode frequency variations are found to be well-correlated with the number of sunspots (gray curve) during cycles 23 and 24 \citep[see also, e.g.,][]{Jain2012}. However, the photometric variability does not exhibit a strong correlation with sunspot number, suggesting that seismic data and photometric variability are independent diagnostics of solar activity. Combining these two independent measures may provide useful insights into the number and spatial distribution of active regions on the surface.

Several activity proxies, including the sunspot numbers and areas, the $10.7$~cm radio flux, and the coronal index, exhibit variations on timescales ranging from $0.6$ to $4$ years, in addition to the well-known 11 year cycle \citep[for a review, see][]{Bazilevskaya2014}. The origin of these variations remains unclear. Of particular interest is the quasi-biennial variation observed in the p-mode frequency shifts, which is also evident in the sunspot numbers and areas, the $10.7$~cm radio flux, and the coronal index \citep{Broomhall2015}. 
To isolate the quasi-biennial variations from the three datasets, we apply a Gaussian filter centered around the period of  $2.5$~yr with a width of $1$~yr (see Figure~\ref{fig:Figure_12}). The correlation coefficient between the filtered travel times ($-\tau$) and the p-mode frequency shifts is found to be significant at $0.88$. 
However, over the period of 1996\,--\,2018, the correlation between the seismic data and the photometric variability is close to zero. This is consistent with previous work by \cite{Broomhall2009}, who also observed differences in the phase between the seismic data and the $10.7$~cm flux.

\begin{figure*}
\includegraphics[width=1.0\textwidth]{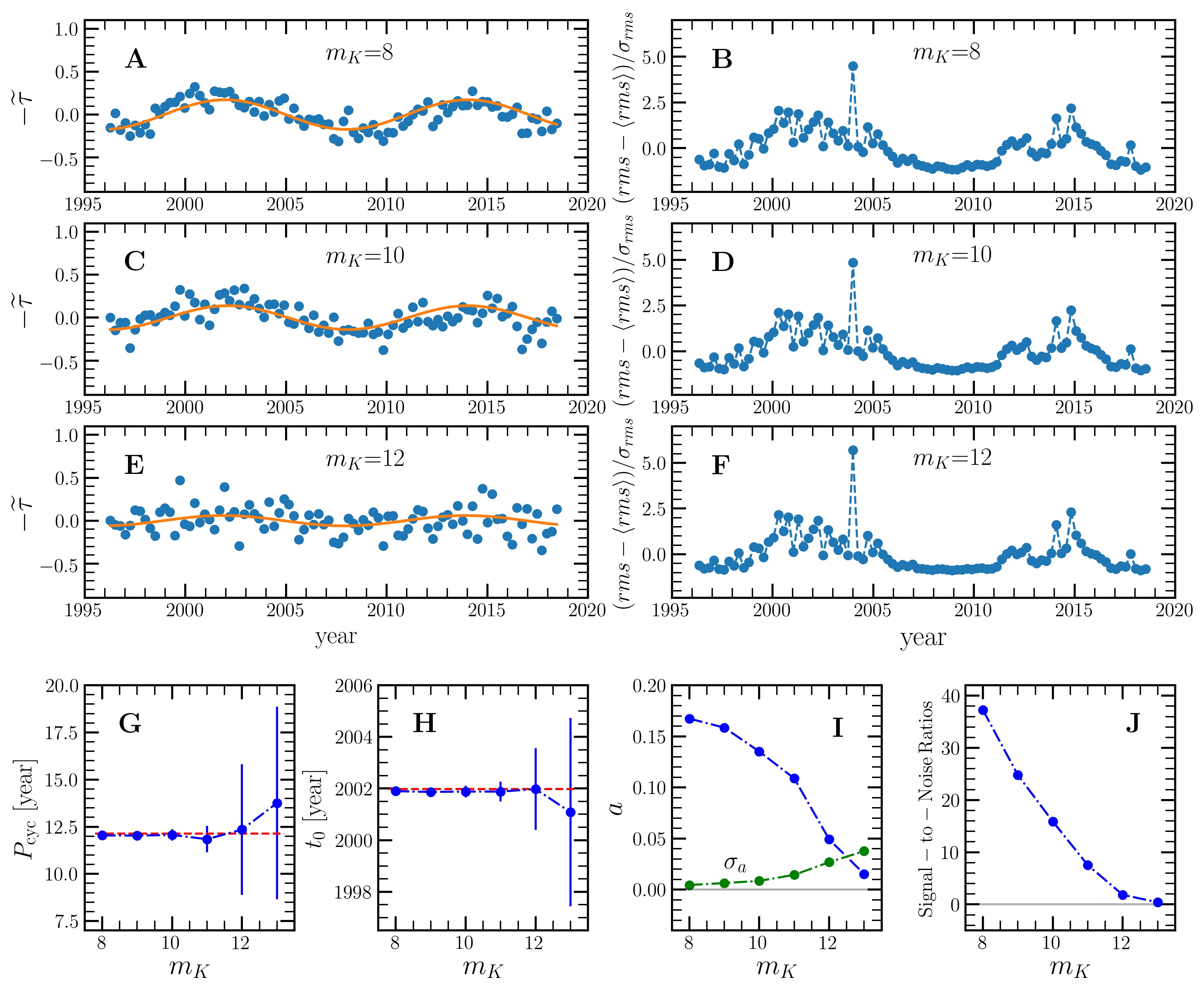}
    \caption{Detectability of the solar cycle in the average travel times in a case of Sun-as-a-Kepler-star observations.
    \textbf{(A, C, E)} Simulated travel times for distant solar analogs obtained by adding photon noise to the VIRGO observations. The apparent  \textit{Kepler} magnitude $m_K$ is indicated on each panel. 
    \textbf{(B, D, F)}  Photometric variability ($RMS$) from the same simulated time series. 
    \textbf{(G, H, I)}  Cycle parameters extracted from fits to the  travel times (orange curves in panels A, C, E), $\widetilde{\tau}_{\mathrm{fit}} (t)  = - a \cos \left(  2\pi(t - t_{0})  /P_\mathrm{cyc} \right)$.
    The means and the error bars are estimated from 20 independent realizations. The green curve shows the standard deviation of the cycle amplitude $a$,  denoted by $\sigma_a$.
   \textbf{(J)} S/N   for the activity cycle amplitude, $a/\sigma_a$, versus \textit{Kepler} magnitude.  }
    \label{fig:Figure_14}
\end{figure*}

\subsection{Prospects for detecting stellar activity cycles} \label{sect:kepler_and_plato}
In this section, we discuss the prospects for applying our method to the analysis of
short cadence stellar observations. 
Here, we only consider distant stars with a magnetic cycle like that of the  Sun and with a rotation axis perpendicular to the line of sight. Different geometrical and magnetic configurations \citep[see, e.g.][]{Gizon2002AN,Papini2019} are beyond the scope of this study. 

In order to simulate broadband photometric data for a Sun-like star, we use the VIRGO/PMO6 total solar irradiance data and introduce random noise to emulate additional shot noise. The PMO6 data span from February 1996 to October 2018 with a temporal cadence of 1 minute. 
 We introduce noise such that the S/N  is Kepler-like for stars of given magnitudes, from  $m_{K}=8$ to $13$. Figure~\ref{fig:Figure_13} shows example simulations (segments of duration 1 year and 2 hours) for stars with magnitudes of $m_{K}=8$, $10$, and $12$. The monthly brightness variations are caused by transits of active regions (faculae and sunspots) that are visible to at least $m_{K}=12$.
On short timescales, the p-mode oscillations are clearly visible up to a magnitude of $m_{K}\approx 10$. 

We cut the light curves into 90 day nonintersecting segments and  measure the p-mode travel times for the first 40 skips.  We then  compute the  average:
\begin{equation}
\langle \tau_s \rangle = \frac{1}{M} \sum_{i=0}^{M-1} \tau_s(t_i)
,\end{equation}
and, for the sake of simplicity, we define the noise with respect to this average:
\begin{equation}
\widetilde{n}_s (t_i) \approx \tau_s(t_i) - \langle \tau_s \rangle.
\end{equation}

Using the noise covariance matrix
\begin{equation}
\widetilde{\Lambda}_{ss'} = \frac{1}{M-1} \sum_{i=0}^{M-1} \widetilde{n}_s (t_i) \widetilde{n}_{s'} (t_i) ,
\end{equation}
we average the travel times  over $N_\mathrm{skip}$:
\begin{equation}
\widetilde{\tau}(t_i) =  \sum_{s=2}^{N_\mathrm{skip}+1} \widetilde{\alpha}_s \, \tau_s(t_i),
\end{equation}
with 
\begin{equation}\widetilde{\alpha}_s = \frac{1}{N_\mathrm{skip}} \sum_{s'=2}^{N_\mathrm{skip}+1}   \Big({\widetilde{\Lambda}}^{-1/2} \Big) _{ss'}.
\end{equation}
We then extract the ``stellar cycle'' from the average travel time  by fitting a simple cosine function,
\begin{equation} \label{eq:model_stars}
\widetilde{\tau}_{\mathrm{fit}} (t)  = - a \cos \left[ 2\pi (t - t_{0}) / P_\mathrm{cyc} \right],
\end{equation}
where $a$, $P_\mathrm{cyc}$, and $t_{0}$ are parameters. 

We estimate the photometric variability of each data segment by calculating the $RMS$ of the flux. In order to determine the amplitude of the activity cycle, we fit a cosine function as described previously. 
Figure~\ref{fig:Figure_14}A-F shows the activity cycle observed in seismic travel times (left panels) and photometric variability (right panels) for apparent magnitudes of $m_{K}=8$, $10$, and $12$. The activity cycle is detected using both methods for $m_{K}=8$ and $10$, but not for $m_{K}=12$, where shot noise is too high for seismic measurements.

To identify the limiting visual magnitude at which the activity cycle is detectable, we define $S/N$ for each magnitude as:
\begin{equation}
S/N (m_K)  = \frac{\mathbb{E} [a(m_K)] }{ \sigma_a (m_K)} ,  
\end{equation}
where $\mathbb{E}[a]$ is the expectation value of the cycle amplitude  and $\sigma_a$ the associated noise level, both obtained from 20 independent realizations of photometric noise (see Figure~\ref{fig:Figure_14}I). In Figure~\ref{fig:Figure_14}J, we find that $S/N \sim 1.5$ for  $m_{K} = 12$.
 A clean cycle detection is achieved for $m_K = 11$, where $S/N\approx8$. For $m_K = 11$, the other two parameters ($P_\mathrm{cyc}$, $t_0$) are also determined to be approximately 8 months and 6 months, respectively,  with high precision. We therefore conclude that the seismic travel-time method is a valuable tool for detecting activity cycles in Sun-like stars.

\begin{figure*}
    \centering
    \includegraphics[width=1.0\textwidth]{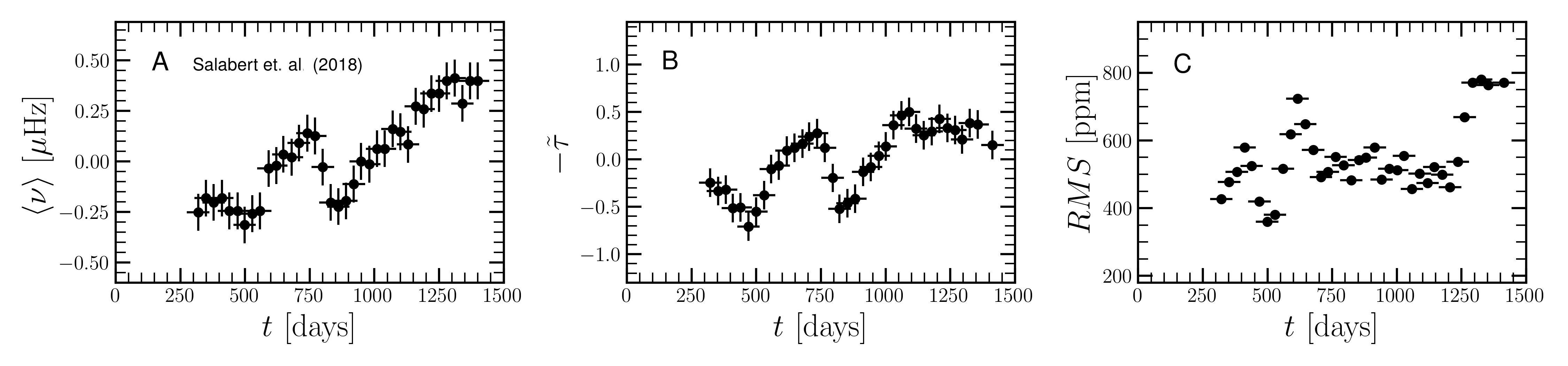}
    \caption{ Variability of the Kepler star HD~173701 (KIC~8006161)
    as inferred from \textbf{(A)} p-mode frequency shifts by \citet{Salabert2018}, \textbf{(B)} average seismic travel-time shift from our work and \textbf{(C)} photometric variability. The vertical errors bars correspond to $\pm 1$ standard deviations. The horizontal error bars represent the observation duration for each measurement (90 days).
  } 
    \label{fig:Figure_15}
\end{figure*}
\subsection{The case of HD~173701}
In order to further illustrate this point, we examine the bright star HD~173701 (KIC~8006161) observed by the \textit{Kepler} mission, with a magnitude of $m_{K}=7.4$. Previous studies by \citet{Kiefer2017} and \citet{Salabert2018} reported strong activity variations in the p-mode frequency shifts during the years 2010\,--\,2014 (see Fig.\ref{fig:Figure_15}A).  Additionally, \citet{Karoff2018} used spectroscopic measurements of the chromospheric emission from multiple epochs spanning from 1979 to 2015  and determined that the star has an activity cycle with a period of $7.4$~yr; during the Kepler observations, it was in the rising phase of this cycle (see their figure 5).

Here, we analyze the star HD~173701 using the methods discussed above and compare with the results from \cite{Salabert2018}.
We divided the data collected by Kepler into segments of  $T=90$~days in length with an overlap of $45$~days. In each segment, we measured the travel times and computed the photometric variability. The seismic activity variations inferred from the travel-time measurements and photometric variability are shown in Fig.~\ref{fig:Figure_15}. Our measurements (Fig.~\ref{fig:Figure_15}B) and those of \citet{Salabert2018} (Fig.~\ref{fig:Figure_15}A) show similar temporal variations over the period of Kepler days from $250$ to $1150$  (see Figs.~\ref{fig:Figure_15}A and B). This includes minima of activity at $\sim 500$ and $800$ days, a local maximum of activity at $750$ days, as well as a rising phase from $800$ to $1100$ days. Beyond approximately $1150$ days, the two datasets are noticably different: while the \citet{Salabert2018} data show an increase in activity, our data reach a plateau.
In Fig.~\ref{fig:Figure_15}C, we show the evolution of the photometric variability, which differs from the seismic data, except for a common minimum of activity at 500 days. We conclude that the two seismic methods give comparable results with similar error bars, while the photometric variability provides an independent diagnostic of stellar activity (as in the case of the Sun).

\section{Conclusion}
\label{sec:conclusions}

    In this work, we describe a new method specifically designed to detect temporal changes in seismic data due to stellar activity, which is simpler than fitting the frequency splittings \citep[cf.][]{Gizon2002AN, Chaplin2003, Benomar2023}.
    Our method works for the Sun and for the bright Kepler star HD~173701. We measured multiple-skip travel times using a cross-correlation technique originally developed  in local helioseismology. 
    Surprisingly, we find that the signature of the solar cycle is strongest in odd skips  from  17 to 31 (i.e., at time lags in the range of 35--64~hours).  

However, we note that the present method focuses only on the asphericity measurements and does not provide any information on other important stellar parameters, such as  stellar rotation rate or inclination angle. Thus, in its current form, our method is no substitute for the peak bagging method.
Despite this limitation,  we expect the method to be useful for analyzing the seismic data from stars to be observed by the upcoming PLATO mission \citep{PLATO2}. 

\begin{acknowledgements}
LG provided the basic idea, VV performed the data analysis, and the authors wrote the paper together.
We thank Rachel Howe for providing the solar p-mode frequency shifts, Wolfgang Finsterle  for the calibrated VIRGO/PMO6 observations, and Damien Fournier for suggesting the whitening transformation. LG acknowledges useful discussions with Othman Benomar.
VV and LG received funding from the Max Planck Society under the grant ``Preparation for PLATO Science'' and from the German Aerospace Center under the grants ``PLATO Data Center''   (50OO1501 and 50OP1902).  This work was supported in part by the ERC Synergy Grant WHOLE SUN 810218 to LG.    
      The VIRGO instrument onboard SoHO is a cooperative effort of scientists, engineers, and technicians to whom we are indebted. SoHO is a project of international collaboration between ESA and NASA. We  used the  open-source  codes \textsc{matplotlib} \citep{2007CSE.....9...90H}, \textsc{numpy}                \citep{5725236}, and \textsc{scipy} \citep{scipy}.
\end{acknowledgements}

\bibliographystyle{aa} 
\bibliography{main}
\end{document}